\documentclass[10pt,conference]{IEEEtran}
\usepackage{bm,cite,graphicx,amssymb,amsmath,color,textcomp}
\usepackage{extarrows,multirow,multicol}
\usepackage{bm}
\usepackage{subeqnarray}
\usepackage{float}
\usepackage{subfig}
\usepackage[noend]{algorithmic}
\usepackage{algorithm}

\newtheorem{remark}{\underline{Remark}}
\newtheorem{lemma}{Lemma}

\newtheorem{theorem}{Theorem}

\newlength{\figwidth}
\IEEEoverridecommandlockouts

\begin{document}
\title{ 
Standard Condition Number–Based Robust\\Signal Detection with Whitening under Uncertainty
\vspace{-4.5mm}
}
\author{
 \IEEEauthorblockN{Tharindu Udupitiya\IEEEauthorrefmark{1}, 
  Saman Atapattu\IEEEauthorrefmark{1}, 
 Prathap Dharmawansa\IEEEauthorrefmark{2}, 
 Chintha Tellambura\IEEEauthorrefmark{3},~and M\'erouane~Debbah\IEEEauthorrefmark{4} 
 }
 \IEEEauthorblockA{
\IEEEauthorrefmark{1}Department of Electrical and Electronic Engineering, School of Engineering, RMIT University, Melbourne, Australia\\
\IEEEauthorrefmark{2}Centre for Wireless Communications,
University of Oulu, Finland\\
\IEEEauthorrefmark{3}Department of Electrical Engineering, University of Alberta, Edmonton, Canada \\
\IEEEauthorrefmark{4}KU 6G Research Center, Khalifa University of Science and Technology, Abu Dhabi, UAE
\vspace{-3mm}}
\vspace{-7.8mm}
}
\maketitle
\begin{abstract}
Robust signal detection in colored noise with unknown covariance is essential in radar, cognitive radio, integrated sensing and communication (ISAC), and quantum sensing applications.  
This paper develops a unified analytical framework for the \textit{Standard Condition Number} (SCN) detector, which employs the ratio of the largest to smallest eigenvalues of the whitened sample covariance matrix.  
The framework jointly covers both \textit{ideal conditions} in which the training and sensing noise statistics are identical and \textit{disturbed conditions} in which interference or jamming alters the sensing covariance.  
Despite the SCN’s practical relevance, its finite-sample false-alarm and detection behavior has not been analytically characterized.  
Using random matrix theory (RMT), we derive general expressions for these probabilities, provide closed-form results for special cases, and show that the SCN preserves the \textit{Constant False Alarm Rate} (CFAR) property under covariance mismatch.  
Analytical and simulation results confirm that the proposed unified framework delivers consistent detection performance and greater robustness than conventional eigenvalue- and LRT-based detectors.

\end{abstract}

\begin{IEEEkeywords}
Standard condition number, eigenvalue detection, noise uncertainty, whitening, random matrix theory.
\end{IEEEkeywords}
\vspace{-2.5mm}

\section{Introduction}\label{S1}
\vspace{-1mm}
Signal detection is fundamental to wireless systems such as radar, cognitive radio (CR), integrated sensing and communications (ISAC), and quantum sensing which underpin emerging 5G and 6G networks~\cite{liu2022integrated, 11126890,Obando25vtc}.  
Practical environments often involve noise uncertainty, interference, and jamming, motivating robust detection strategies such as the likelihood ratio test (LRT), generalized LRT (GLRT), energy detection (ED), and eigenvalue-based methods~\cite{atapattu2011energy}.  
Among them, eigenvalue-based detection is particularly effective in low-SNR or model-mismatch conditions by exploiting the statistical structure of the sample covariance matrix~\cite{7903598,Chamain,debbah2011,10437271}.  
These detectors form test statistics from the maximum, minimum, or combined eigenvalues, balancing robustness and complexity~\cite{johnstoneRoy,dissanayake2022}.  
In this context, the \textit{standard condition number} (SCN), defined as the ratio of the largest to smallest eigenvalue, stands out for its inherent \emph{scale-invariance}~\cite{5089517,matthaiou2010,zhang2017}, enabling CFAR behavior under noise and interference uncertainty.  

Detectors such as the GLRT and ED assume perfectly known noise statistics.  
In practice, wireless channels exhibit colored noise arising from interference, multipath, and hardware imperfections, leading to spatial and temporal correlation~\cite{10437271,Chamain,besson2023}.  
Such mismatches degrade detection reliability.  
A practical remedy is to estimate the noise covariance from \textit{noise-only} (secondary) samples and apply \textit{signal whitening}~\cite{Chamain,besson2023,besson2016generalized,werner2007,johnstone2020,10619207}, transforming correlated noise into an approximately white process.  
This work employs the SCN detector within this whitening framework to achieve robust, analytically tractable performance under realistic conditions.

Detection \emph{without whitening} has been widely studied for detectors such as the ED~\cite{atapattu2011energy}, GLRT~\cite{debbah2011,mcwhorter2023}, and several eigenvalue-based methods including the maximum eigenvalue test~\cite{johnstoneRoy,debbah2011,Dissanayake25vtc,Obando25vtc}, trace-to-maximum ratio tests~\cite{dissanayake2022,5089517}, and the SCN~\cite{kobeissi2017approximating,matthaiou2010,zhang2017,nafkha2020standard}.  
In contrast, \emph{whitening-based detection} has been examined mainly for the GLRT~\cite{besson2023,besson2016generalized}, LRT~\cite{werner2007,johnstone2020}, and maximum-eigenvalue detectors~\cite{Chamain, 10619207}, typically under restrictive noise models.  
However, the incorporation of whitening into SCN-based detection remains largely unexplored.  
Moreover, existing whitening-based schemes assume that noise statistics estimated from secondary samples remain valid during detection on primary (signal-plus-noise) data~\cite{besson2016generalized,Chamain,werner2007}.  
In practice, dynamic factors such as jamming or active interference can alter the sensing environment, leading to covariance mismatch and degraded reliability, which this work explicitly addresses as a limitation.

Despite extensive studies on eigenvalue-based detectors, SCN detection lacks a rigorous analytical treatment even under ideal conditions and has not been examined under noise uncertainty or whitening.  
This paper addresses these gaps by \textit{developing the first unified analytical framework that jointly models both ideal and disturbed sensing conditions}, thereby covering scenarios in which the noise covariance during detection may differ from that estimated in training due to interference or jamming.  
The main contributions are as follows:  
(i) using orthogonal polynomial methods from random matrix theory (RMT)~\cite{Chamain,dissanayake2022}, we derive analytical expressions for the SCN false-alarm and detection probabilities under general parameters and provide closed-form results for special cases (Sec.~III);  
(ii) we extend the framework to non-stationary settings with covariance mismatch and show that the SCN detector preserves CFAR behavior (Sec.~IV); and  
(iii) analytical and simulation results confirm that SCN achieves robust detection performance, outperforming LRT and max-eigenvalue in terms of CFAR and total error (Sec.~V).

\begin{figure*}[t!]
    \centering
    \subfloat[Training phase: noise-only samples for whitening.\label{fig:fig1}]{\includegraphics[width=0.3305\textwidth]{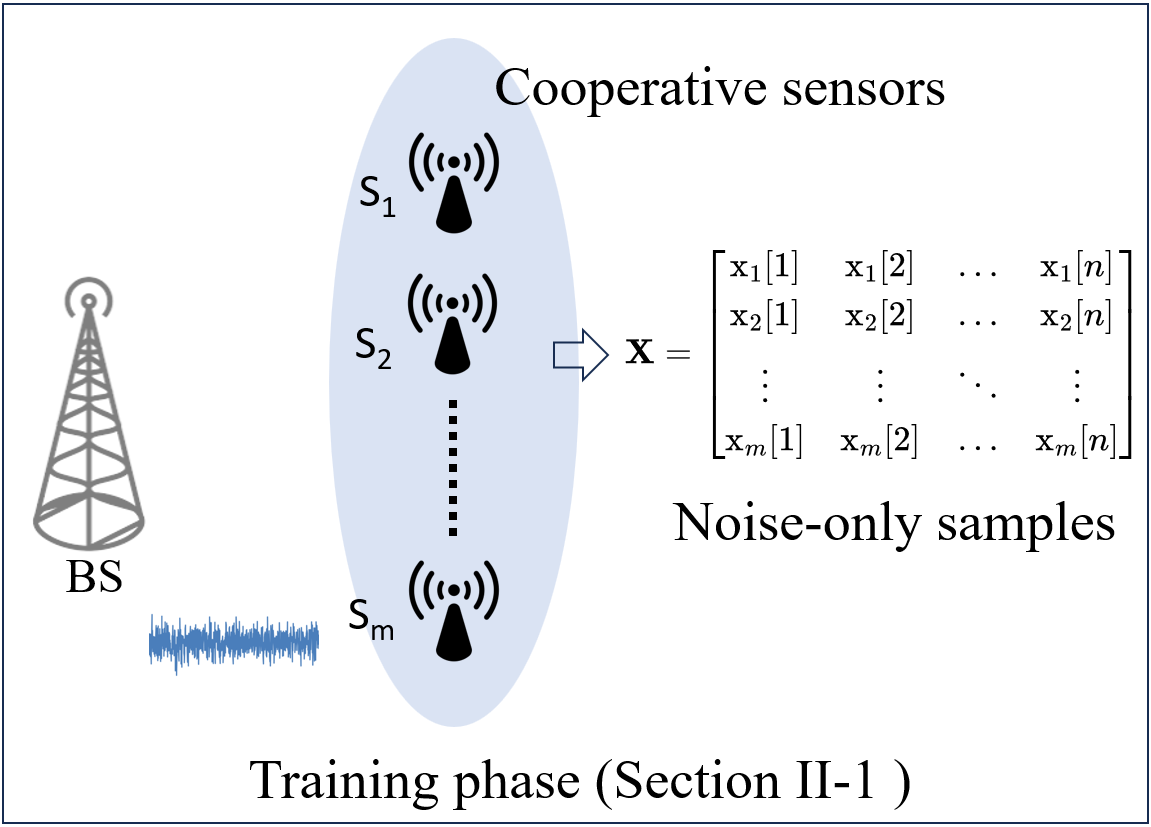}} 
    \hfill
    \subfloat[Sensing phase: signal-plus-noise samples.\label{fig:fig2}]{\includegraphics[width=0.332\textwidth]{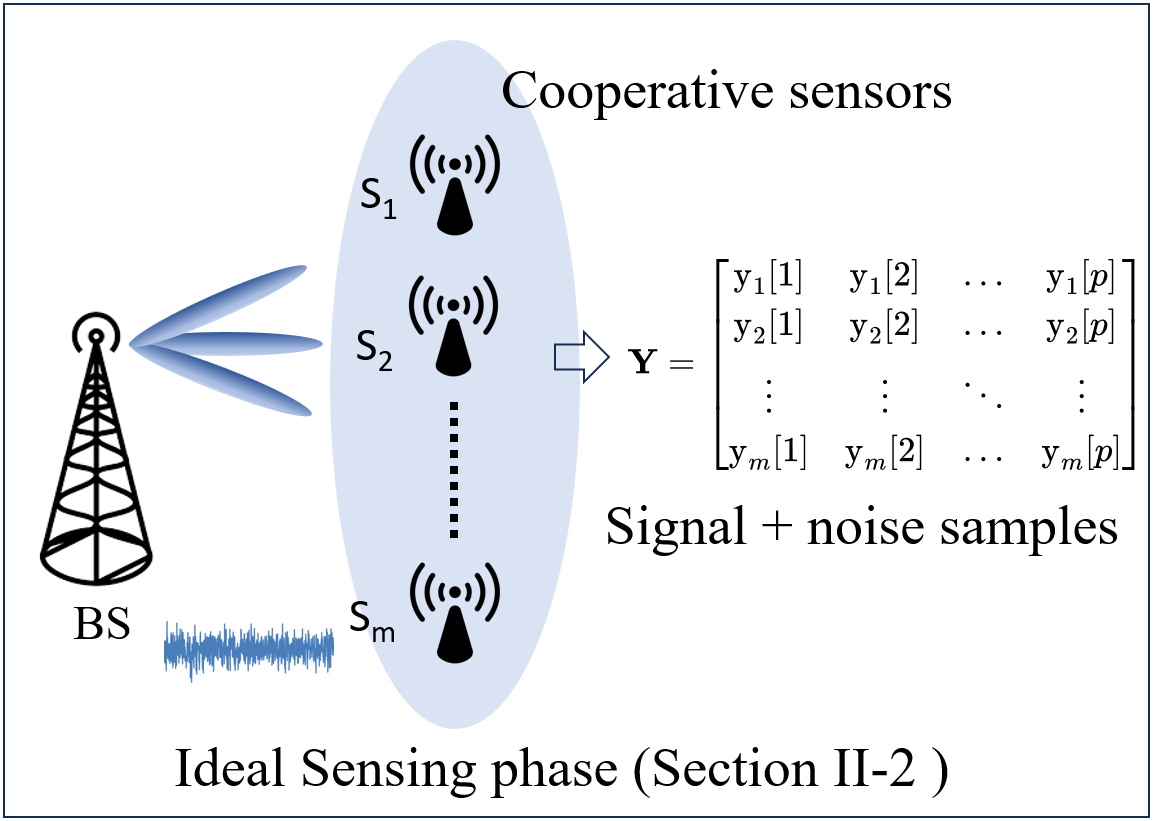}} 
    \hfill
    \subfloat[Disturbed Sensing Phase: interference at sensing.\label{fig:fig3}]{\includegraphics[width=0.333\textwidth]{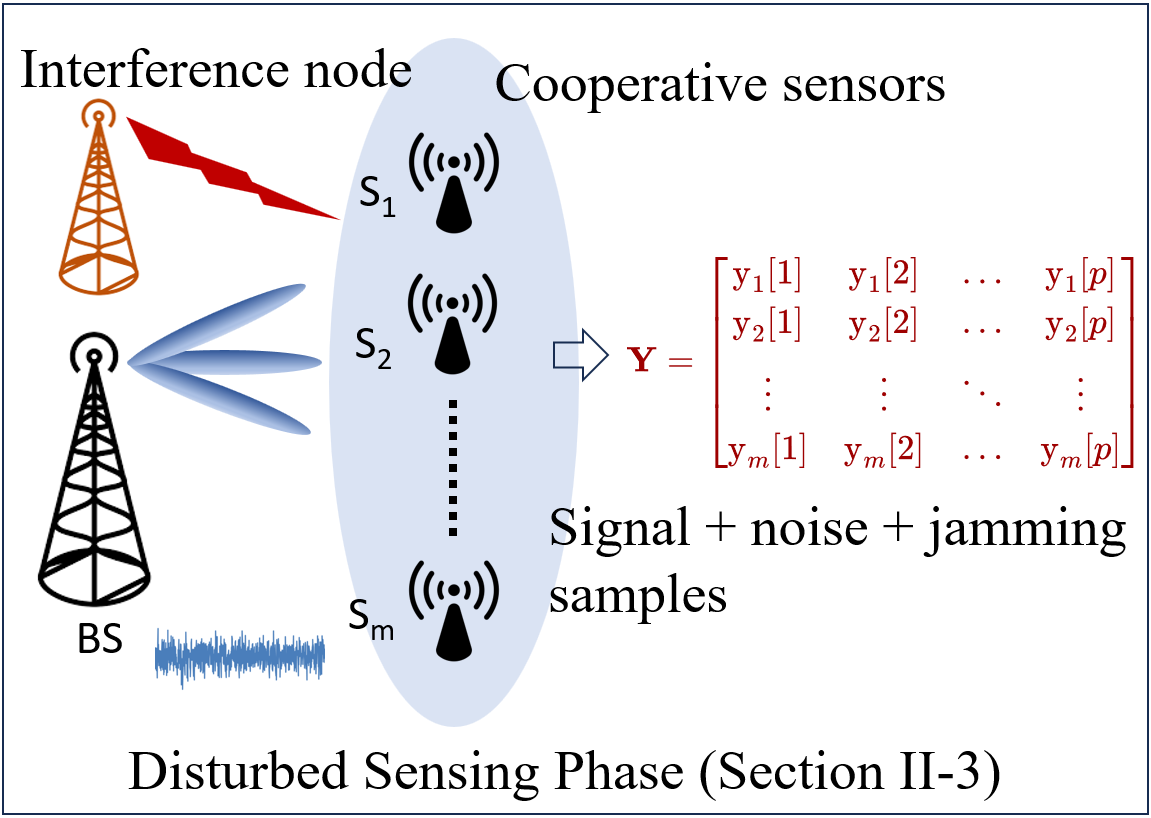}} 
    \caption{System model illustrating the two-phase training and sensing for the detection process and the impact of jamming.}
    \label{fig:system_model}\vspace{-5mm}
\end{figure*}

\section{System Model and Signal Model}
We consider a cooperative wireless sensing network in which a base station (BS) transmits a signal to be detected by \( m \) distributed single-antenna sensors \( { \textsf{S}_1, \dots, \textsf{S}_m } \) (Fig.~\ref{fig:system_model}).
The detection process operates in two stages: a \emph{training phase} and a \emph{sensing phase}. During the \textit{training phase} (Fig.~\ref{fig:fig1}), the BS remains silent, and each sensor collects ambient noise samples used to estimate the spatial noise covariance and construct a whitening matrix.
In the subsequent \textit{sensing phase} (Fig.~\ref{fig:fig2}), the sensors observe signal-plus-noise snapshots for hypothesis testing.
Ideally, the noise statistics remain identical to those estimated during training; however, in practical deployments, unintended interference or deliberate jamming (Fig.~\ref{fig:fig3}) can alter the noise power or correlation structure.
This mismatch between training and sensing covariances, referred to as \emph{noise uncertainty}, directly impacts the robustness and detection behavior of classical detectors. 

\subsubsection{Training Phase (Fig.~\ref{fig:fig1})}
During the silent training period, the BS remains inactive, and each sensor collects $n$ noise-only snapshots modeled for \(\ell = 1, \ldots, n\) as
\begin{equation}\label{eq_H0H1a}
    \mathbf{x}[\ell] = \mathbf{n}[\ell] \text{ and } \mathbf{X} = [\mathbf{x}[1], \dots, \mathbf{x}[n]] \in \mathbb{C}^{m \times n},
\end{equation}
where $\mathbf{x}[\ell] \in \mathbb{C}^{m}$ denotes the received noise vector across the $m$ sensors and $\mathbf{n}[\ell] \sim \mathcal{CN}_m(\mathbf{0}, \boldsymbol{\Sigma})$ with unknown spatial covariance $\boldsymbol{\Sigma} \in \mathbb{C}^{m \times m}$.  

\subsubsection{Ideal Sensing Phase (Fig.~\ref{fig:fig2})}
During sensing, each sensor acquires $p$ signal-plus-noise snapshots modeled by the binary hypothesis test
\begin{equation}\label{eq_H0H1}
\mathbf{y}_{\mathrm{I}}[k] =
\begin{cases}
\mathbf{n}[k], & \mathcal{H}_0,\\
\sqrt{\rho}\,\mathbf{h}s[k] + \mathbf{n}[k], & \mathcal{H}_1,
\end{cases}
\quad k = 1,\dots,p,
\end{equation}
where $\mathbf{y}_{\mathrm{I}}[k] \in \mathbb{C}^m$ is the received vector across $m$ sensors, 
$s[k] \sim \mathcal{CN}(0,1)$ denotes the transmitted symbol with power $\rho$, and 
$\mathbf{h} \in \mathbb{C}^m$ represents the deterministic BS–sensor channel.  
The additive noise $\mathbf{n}[k] \sim \mathcal{CN}_m(\mathbf{0}, \boldsymbol{\Sigma})$ shares the same covariance structure as in training.  

\subsubsection{Disturbed Sensing Phase (Fig.~\ref{fig:fig3})}\label{subsec:disturbed}
In practical environments, the sensing phase is often affected by additional disturbances such as unintended interference or deliberate jamming, leading to a mismatch between the training and sensing noise statistics, as illustrated in Fig.~\ref{fig:fig3}.  
Let $\mathbf{i}[k] \in \mathbb{C}^{m}$ denote the disturbance component observed during sensing.  
The received signal is modeled as
\begin{equation}\label{eq_H0H1_jam}
\mathbf{y}_{\mathrm{d}}[k] =
\begin{cases}
\mathbf{n}[k] + \mathbf{i}[k], & \mathcal{H}_0, \\[3pt]
\sqrt{\rho}\,\mathbf{h}s[k] + \mathbf{n}[k] + \mathbf{i}[k], & \mathcal{H}_1,
\end{cases}
\end{equation}
where $\mathbf{n}[k] \sim \mathcal{CN}_m(\mathbf{0}, \boldsymbol{\Sigma})$ denotes the receiver noise estimated during training, and $\mathbf{i}[k] \sim \mathcal{CN}_m(\mathbf{0}, \boldsymbol{\Sigma}_i)$ represents colored Gaussian disturbance. The disturbance covariance is modeled as a scaled version of the nominal noise covariance, $\boldsymbol{\Sigma}_i = \mu'\boldsymbol{\Sigma}$, resulting in an effective sensing-phase covariance $\mu\boldsymbol{\Sigma}$, where $\mu = 1 + \mu'$ denotes the \textit{covariance mismatch factor}, quantifying the deviation between training and sensing noise.

Arranging all training snapshots yields $\mathbf{X} \in \mathbb{C}^{m \times n}$ from \eqref{eq_H0H1a}, which is used to estimate the noise covariance $\boldsymbol{\Sigma}$ and construct the whitening matrix.  
Similarly, by arranging all sensing snapshots from \eqref{eq_H0H1} or \eqref{eq_H0H1_jam} forms $\mathbf{Y}_{\mathrm{I}}$ or $\mathbf{Y}_{\mathrm{d}}$, respectively, where $\mathbf{Y} = [\mathbf{y}[1], \dots, \mathbf{y}[p]] \in \mathbb{C}^{m \times p}$, used in subsequently. 

\subsubsection{Signal Whitening Unified Model}\label{sec:whitening}
In multi-antenna sensing systems, spatial correlation among sensor noises arises from hardware coupling or external interference.  
To suppress this effect, a whitening transformation is applied using the covariance estimated during the training phase.  
Let $\mathbf{S}_{\mathrm{I}}$ and $\mathbf{S}_{\mathrm{d}}$ denote the true covariances of the received data during the ideal and disturbed sensing phases, respectively:
\[
\mathbf{S}_{\mathrm{I}} =
\begin{cases}
\boldsymbol{\Sigma}, & \mathcal{H}_0,\\[2pt]
\rho\,\mathbf{h}\mathbf{h}^\dagger + \boldsymbol{\Sigma}, & \mathcal{H}_1,
\end{cases}
\qquad
\mathbf{S}_{\mathrm{d}} =
\begin{cases}
\mu\boldsymbol{\Sigma}, & \mathcal{H}_0,\\[2pt]
\rho\,\mathbf{h}\mathbf{h}^\dagger + \mu\boldsymbol{\Sigma}, & \mathcal{H}_1,
\end{cases}
\]
where $\mu > 0$ characterizes the level of covariance mismatch between training and sensing due to interference or jamming.  
The ideal case is recovered by setting $\mu = 1$.  
Accordingly, all subsequent formulations are developed for the general disturbed case, with $\mu$ serving as an auxiliary parameter that reduces to the ideal setting when $\mu=1$.

Let $\mathbf{S} \in \{\mathbf{S}_{\mathrm{I}},\mathbf{S}_{\mathrm{d}}\}$.  
Corresponding whitened covariance is
\begin{equation}
\bm{\Psi} = \boldsymbol{\Sigma}^{-1}\mathbf{S} =
\begin{cases}
\mu\mathbf{I}_m, & \mathcal{H}_0,\\
\rho\,\boldsymbol{\Sigma}^{-1}\mathbf{h}\mathbf{h}^\dagger + \mu\mathbf{I}_m, & \mathcal{H}_1. 
\end{cases}\label{eq:whitened_covariance}
\end{equation}
 
\begin{lemma}[Rank-One Perturbation]
Under $\mathcal{H}_1$, the whitened covariance
\(
\bm{\Psi}=\mu\mathbf{I}_m+\rho\,\boldsymbol{\Sigma}^{-1}\mathbf{h}\mathbf{h}^\dagger
\)
differs from $\mu\mathbf{I}_m$ by a rank-one matrix, i.e.,
\(
\operatorname{rank}\!\big(\bm{\Psi}-\mu\mathbf{I}_m\big)=1
\)
whenever $\mathbf{h}\neq \mathbf{0}$.
\end{lemma}

\begin{IEEEproof}
Set $\mathbf{A}\triangleq \bm{\Psi}-\mu\mathbf{I}_m=\rho\,\boldsymbol{\Sigma}^{-1}\mathbf{h}\mathbf{h}^\dagger$.
Since $\rho>0$ is a scalar, $\operatorname{rank}(\mathbf{A})=\operatorname{rank}\!\big(\boldsymbol{\Sigma}^{-1}\mathbf{h}\mathbf{h}^\dagger\big)$.
Because $\boldsymbol{\Sigma}\succ 0$, $\boldsymbol{\Sigma}^{-1}$ is invertible, hence
\(
\operatorname{rank}\!\big(\boldsymbol{\Sigma}^{-1}\mathbf{h}\mathbf{h}^\dagger\big)
=\operatorname{rank}\!\big(\mathbf{h}\mathbf{h}^\dagger\big)=1
\)
when $\mathbf{h}\neq \mathbf{0}$ (outer products of nonzero vectors have rank one).  
Equivalently, note that $\mathbf{A}\neq \mathbf{0}$ since
\(
\mathbf{A}\mathbf{h}=\rho\,\boldsymbol{\Sigma}^{-1}\mathbf{h}\,(\mathbf{h}^\dagger\mathbf{h})\neq \mathbf{0},
\)
and $\operatorname{rank}(\mathbf{A})\le 1$ by the submultiplicativity
\(
\operatorname{rank}(\mathbf{A})=\operatorname{rank}(\boldsymbol{\Sigma}^{-1}\mathbf{h}\mathbf{h}^\dagger)\le
\min\{\operatorname{rank}(\boldsymbol{\Sigma}^{-1}),\operatorname{rank}(\mathbf{h}\mathbf{h}^\dagger)\}=1.
\)
Thus $\operatorname{rank}(\mathbf{A})=1$.
\end{IEEEproof}

\begin{remark}[Eigenstructure (useful for thresholds)]
By congruence with $\boldsymbol{\Sigma}^{1/2}$,
\(
\boldsymbol{\Sigma}^{1/2}\bm{\Psi}\boldsymbol{\Sigma}^{-1/2}
=\mu\mathbf{I}_m+\rho\,\mathbf{u}\mathbf{u}^\dagger,
\quad \mathbf{u}\triangleq \boldsymbol{\Sigma}^{-1/2}\mathbf{h}.
\)
Hence $\bm{\Psi}$ has one eigenvalue $\mu+\rho\,\|\mathbf{u}\|^2=\mu+\rho\,\mathbf{h}^\dagger\boldsymbol{\Sigma}^{-1}\mathbf{h}$
and $m\!-\!1$ eigenvalues equal to $\mu$. This confirms that $\mathcal{H}_1$ induces a rank-one spectral inflation over $\mu\mathbf{I}_m$.
\end{remark}

In practice, $\boldsymbol{\Sigma}$ and $\mathbf{S} \in \{\mathbf{S}_{\mathrm{I}},\mathbf{S}_{\mathrm{d}}\}$ are unknown and are estimated from training $\mathbf{X}$ and sensing $\mathbf{Y}\in \{\mathbf{Y}_{\mathrm{I}},\mathbf{Y}_{\mathrm{d}}\}$ data as 
\begin{subequations}
\begin{align}
n\,\widehat{\boldsymbol{\Sigma}} &= \mathbf{X}\mathbf{X}^\dagger = \sum_{\ell=1}^{n}\mathbf{x}[\ell]\mathbf{x}[\ell]^\dagger,\\
p\,\widehat{\mathbf{S}} &= \mathbf{Y}\mathbf{Y}^\dagger = \sum_{k=1}^{p}\mathbf{y}[k]\mathbf{y}[k]^\dagger,
\end{align}
\end{subequations}
with $n,p \ge m$ ensuring invertibility with high probability.  
The resulting sample-whitened covariance is
\(\widehat{\bm{\Psi}} \;\triangleq\; \widehat{\boldsymbol{\Sigma}}^{-1}\widehat{\mathbf{S}},\) 
which coincides with the population-level expression in \eqref{eq:whitened_covariance} and captures whitening mismatch when $\mu\neq1$. This $\widehat{\bm{\Psi}}$ is the basis for the subsequent detection rules.

\subsubsection{Motivation for SCN under Noise Uncertainty}
\begin{figure}[t!]
    \centering
    \includegraphics[width=0.85\columnwidth]{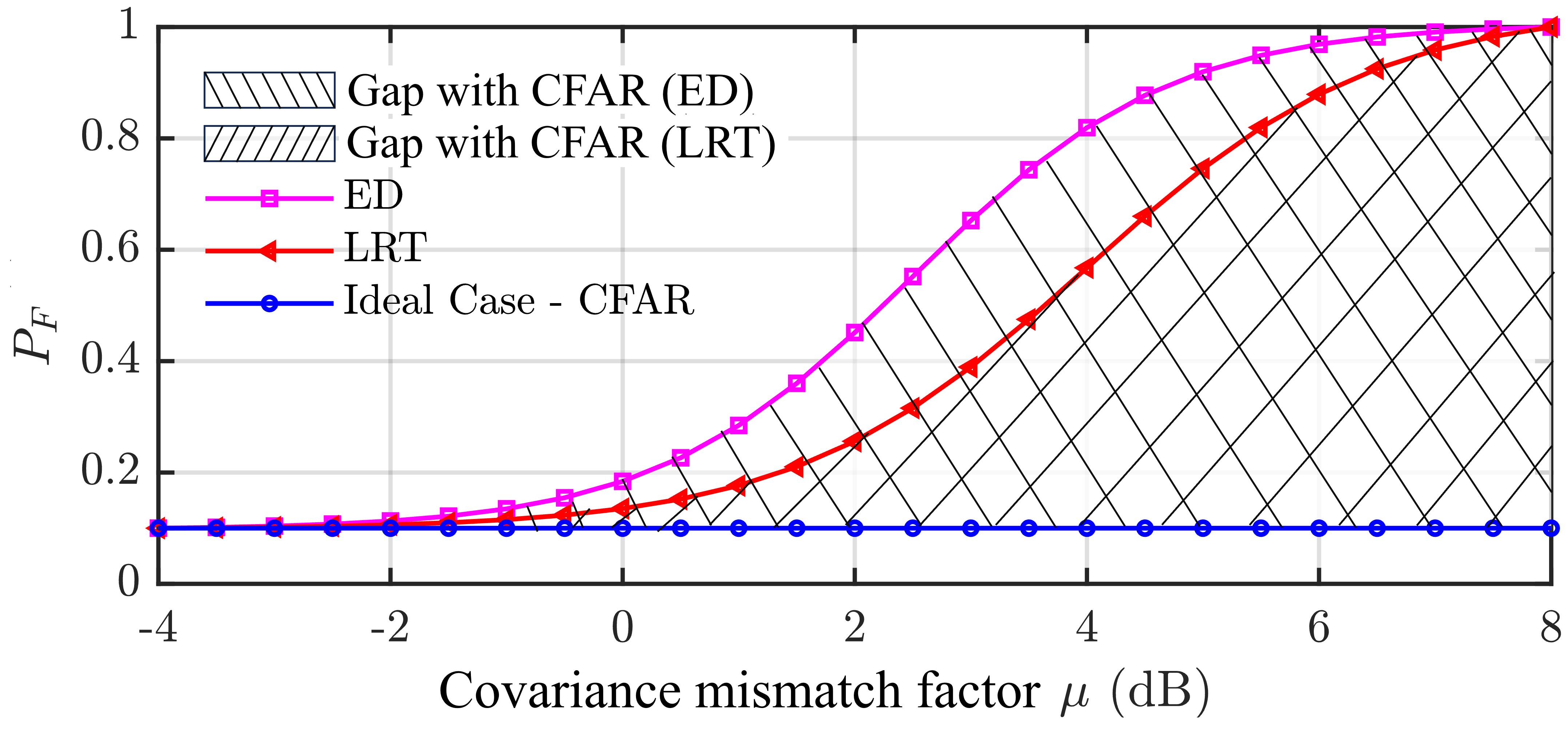}
    \caption{False-alarm probability $P_{\mathrm{F}}$ versus noise-mismatch level $\mu$ showing degradation of LRT and ED under uncertainty.
    }
    \label{fig:lost_spectrum_opportunity}\vspace{-5mm}
\end{figure}

In practice, the sensing system is unaware of whether the environment remains ideal or becomes disturbed.  
Hence, conventional detectors such as the LRT, energy detector (ED), or the largest eigenvalue ($\lambda_{\max}$) test are typically designed under the ideal model assuming $\mu=1$.  
When unexpected interference or jamming alters the effective noise covariance ($\mu \neq 1$), these detectors experience mismatched whitening, resulting in biased decision thresholds and degraded CFAR behavior.  
As illustrated in Fig.~\ref{fig:lost_spectrum_opportunity}, such sensitivity to noise-power variation increases the false-alarm rate and causes missed spectrum opportunities for secondary users.

To achieve robustness against these disturbances, the detector must rely on scale-invariant statistics that remain insensitive to covariance mismatches.  
Motivated by this requirement, a ratio-based test statistic known as the \emph{Standard Condition Number} (SCN) is adopted.  
The SCN depends only on the relative spread of the eigenvalues of the whitened covariance matrix, providing a CFAR-robust alternative for reliable detection under interference and noise uncertainty.

\vspace{-1mm}
\section{Standard Condition Number (SCN) Detector}
The SCN detector exploits the eigenvalue spread of the whitened sample covariance matrix to achieve robustness against noise and interference uncertainty~\cite{besson2023,Chamain,besson2016generalized}.  
The test statistic is defined as
\begin{align}\label{eq_scn}
\kappa(\widehat{\boldsymbol{\Psi}}) =
\frac{\lambda_{\max}(\widehat{\boldsymbol{\Psi}})}{\lambda_{\min}(\widehat{\boldsymbol{\Psi}})}, 
\quad \text{where} \quad
\widehat{\boldsymbol{\Psi}} = \widehat{\boldsymbol{\Sigma}}^{-1}\widehat{\mathbf{S}}.
\end{align}

Under the unified disturbed model of Sec.~\ref{subsec:disturbed}, the sensing covariance follows
\begin{equation}\label{eq:wishart_general}
p\,\widehat{\mathbf{S}} \sim
\begin{cases}
\mathcal{CW}_m(p,\, \mu\boldsymbol{\Sigma}), & \mathcal{H}_0,\\[2pt]
\mathcal{CW}_m(p,\, \rho\mathbf{h}\mathbf{h}^\dagger + \mu\boldsymbol{\Sigma}), & \mathcal{H}_1,
\end{cases}
\end{equation}
where $\mu>0$ denotes the covariance–mismatch factor introduced by interference or jamming.  
After whitening by $\widehat{\boldsymbol{\Sigma}}^{-1}$, the distribution becomes
\begin{equation}\label{eq:white_wishart_general}
p\,\widehat{\boldsymbol{\Psi}} \sim
\begin{cases}
\mathcal{CW}_m(p,\, \mu\mathbf{I}_m), & \mathcal{H}_0,\\[2pt]
\mathcal{CW}_m(p,\, \mu\mathbf{I}_m + \gamma\,\mathbf{v}\mathbf{v}^\dagger), & \mathcal{H}_1,
\end{cases}
\end{equation}
where $\mathbf{v} = \boldsymbol{\Sigma}^{-1/2}\mathbf{h}/\|\boldsymbol{\Sigma}^{-1/2}\mathbf{h}\|$ and $\gamma = \rho\,\mathbf{h}^\dagger\boldsymbol{\Sigma}^{-1}\mathbf{h}$ denote the signal direction and the nominal SNR.

\begin{lemma}[CFAR Invariance of the SCN Statistic]
Under $\mathcal{H}_0$, scaling the covariance by $\mu>0$ uniformly scales all eigenvalues of $\widehat{\boldsymbol{\Psi}}$ as
\(
\lambda_i^{(\mu)} = \mu\,\lambda_i^{(1)},\,\, i=1,\ldots,m.
\)
Thus, SCN
\[
\kappa^{(\mu)}(\widehat{\boldsymbol{\Psi}}) 
= \frac{\lambda_{\max}^{(\mu)}}{\lambda_{\min}^{(\mu)}}
= \frac{\lambda_{\max}^{(1)}}{\lambda_{\min}^{(1)}}
= \kappa^{(1)}(\widehat{\boldsymbol{\Psi}}),
\]
is invariant to the scaling factor $\mu$, demonstrating its CFAR property w.r.t. noise-power and covariance mismatches.
\end{lemma}

\begin{remark}[Behavior under $\mathcal{H}_1$]
Under $\mathcal{H}_1$, the sensing covariance becomes $(\mu\mathbf{I}_m + \gamma\mathbf{v}\mathbf{v}^\dagger)$, where $\mu$ scales the background noise while leaving the rank-one signal component unchanged.  
The effective SNR reduces to $\gamma/\mu$: higher $\mu$ (strong interference or jamming) diminishes detection sensitivity, whereas lower $\mu$ improves it.  
Thus, while the SCN detector preserves CFAR robustness under $\mathcal{H}_0$, its detection probability naturally tracks the physical SNR variation under $\mathcal{H}_1$.
\end{remark}

As shown in Fig.~\ref{fig:lost_spectrum_opportunity}, classical detectors such as the LRT, ED, and $\lambda_{\max}$ test suffer rising false-alarm rates as $\mu$ increases, whereas the SCN maintains constant false-alarm, enabling reliable spectrum access under interference and noise uncertainty.

\begin{remark}[Novelty of Analytical Treatment]
To the best of our knowledge, no prior work has provided a rigorous analytical characterization of the SCN detector, even under the ideal case ($\mu=1$).  
The framework developed in this paper generalizes that scenario to the more practical setting of disturbed noise statistics ($\mu\neq1$), enabling unified analysis of both stationary and non-stationary environments.
\end{remark}

\begin{figure*}
\begin{equation}\label{eq:P(a,b,m)}
\begin{split}
K_m(\alpha,\beta)&=\frac{(-1)^{\beta(\alpha+m-1)}2^{\alpha(m+\beta)}\mathcal{K}(m,n,p)}{(m-1)!}\prod_{i=1}^{\alpha}\frac{(m+i)!}{[(i-1)!]^2\,(m+\alpha+\beta-i-1)!}\prod_{j=1}^{\beta}\frac{(m+j)!}{[(j-1)!]^2\,(m+\alpha+\beta-j-1)!}\\
&\times \prod_{i=1}^{\alpha+\beta}\frac{\left[(m-2+i)!\right]^2}{(2m+2i-2)!}\prod_{j=0}^{m-2}\frac{j!(j+1)!(j+2)!}{(m+j+1)!}\det\left[\mathcal{U}_{i,k}(m,\alpha,\beta) \quad  \mathcal{V}_{i,\ell}(m,\alpha,\beta)\right]_{\substack{
i=1,2,\ldots,\alpha + \beta\\
k=1,2,\ldots,\alpha\\
\ell=\alpha+1,\alpha+2,\ldots,\alpha + \beta
}}
\end{split}
\end{equation}
\begin{equation}\label{eq:S(t,y)}
\begin{split}
\mathcal{Q}_{j_k}(t,y)=&\sum_{j_k=0}^{m+\alpha+\beta-k-1}\frac{(m+1+k)_{j_k}}{j_k!(k)_{j_k}}
(-(m+\alpha+\beta-k-1))_{j_k}\left[\frac{-(y+1)}{y(t-1)}\right]^{j_k}\\
\end{split}
\end{equation}
\rule{\textwidth}{0.5pt}
\begin{equation}\label{eq:tildaQ}
\widetilde{K}_m(\beta)=\frac{(-1)^{\beta(m-1)}\mathcal{K}(m,m,p)}{(m-1)!}\prod_{i=1}^{\beta}\frac{[(m-2+i)!]^2}{(2m+2i-2)!}\prod_{j=0}^{m-2}\frac{j!(j+1)!(j+2)!}{(m+j+1)!}\det\left[\mathcal{V}_{i,\ell}(m,0,\beta)\right]_{\substack{
i=1,2,\ldots,\beta\\
\ell=1,2,\ldots, \beta
}}
\end{equation}
\rule{\textwidth}{0.5pt}
\begin{equation}\label{eq:I_A}
\mathcal{I}^A_m(t)= \frac{(-1)^{m-1}m}{\gamma_{\!e}^{m-1}(1+\gamma_{\!e})^{(m-1)^2}}\left(1-\frac{1}{t}\right)^{\widetilde{m}-1}\text{B}(\widetilde{m},\widetilde{m})F_1\left(\widetilde{m};\widetilde{m}-1,(m-1)^2;2\widetilde{m};1-\frac{1}{t(\gamma_{\!e}+1)},\frac{\gamma_{\!e}}{\gamma_{\!e}+1}\right)
\end{equation} 
\begin{equation}\label{eq:I_B}
\begin{split}
\mathcal{I}^B_m(t)&=m\sum_{k=0}^{m-2}\frac{(-1)^k(m+k)!}{k!}\sum_{j=0}^{m-2-k}\frac{(j+1)(t-1)^{\widetilde{m}-1+j}}{(k+2+j)!(m-k-2-j)!}\sum_{\ell=0}^{\infty}\frac{(m-1)_{\ell}}{\ell!}\frac{\gamma_{\!e}^{\ell+j-m+1}}{(1+\gamma_{\!e})^{\ell+j+1}}\\
& \times B(\widetilde{m}+j,\widetilde{m}+j+\ell)F_1\left(\widetilde{m}+j+\ell;\widetilde{m}-3,(j+2);2(\widetilde{m}+j)+\ell;(t-1),\left(1-\frac{t}{\gamma_{\!e}+1}\right)\right)
\end{split}
\end{equation}
\rule{\textwidth}{0.5pt}
\end{figure*}

\section{Detection Performance of the SCN $\kappa(\widehat{\boldsymbol{\Psi}})$}\label{sec:analytical_expression}
We characterize the false-alarm and detection probabilities of the SCN statistic in \eqref{eq_scn}:
\begin{align}
P_{\mathrm{F}}(t) 
&\triangleq \Pr\!\left(\kappa(\widehat{\boldsymbol{\Psi}})>t \mid \mathcal{H}_0\right)
= 1 - F_{\kappa}(t;\mathcal{H}_0), \\
P_{\mathrm{D}}(\gamma,\mu,t) 
&\triangleq \Pr\!\left(\kappa(\widehat{\boldsymbol{\Psi}})>t \mid \mathcal{H}_1\right)
= 1 - F_{\kappa}(t;\mathcal{H}_1),
\end{align}
where $t$ is the decision threshold and $F_{\kappa}(\cdot;\mathcal{H}_i)$ denotes the CDF of $\kappa(\widehat{\boldsymbol{\Psi}})$ under $\mathcal{H}_i$, $i\in\{0,1\}$.



Analytical integral expressions for $F_{\kappa}(t;\mathcal{H}_0)$ and $F_{\kappa}(t;\mathcal{H}_1)$ are provided in Theorem~\ref{th:1} (Eq.~\eqref{eq:cdf_integral}) and Theorem~\ref{th:2} (Eq.~\eqref{eq:H1_integral}), enabling evaluation of $P_{\mathrm{F}}(t)$ and $P_{\mathrm{D}}(\gamma,\mu,t)$.

\begin{theorem}\label{th:1}
Let \( \textbf{W}_1 \sim \mathcal{W}_m(p, \mu\textbf{I}_m) \) and \( \textbf{W}_2 \sim \mathcal{W}_m(n, \textbf{I}_m) \)
be two independent Wishart matrices with \( p, n \geq m \). Then
the c.d.f. of \(\kappa(\bm{\widehat{\Psi}})\) of \(\bm{\widehat{\Psi}}=\mu^{-1} \textbf{W}_1 \textbf{W}_2^{-1} \), for $t>1$, is given by
\end{theorem} 
\begin{equation}\label{eq:cdf_integral}
\begin{split}
&F_{\kappa}(t;\mathcal{H}_0)\\
& =K_m(\alpha,\beta)\left(t-1\right)^{m(\beta+m)-\alpha-\beta-1}\\
& \quad   \times\int_0^\infty\hspace{-1mm}\frac{\prod_{k=1}^{\alpha}\mathcal{Q}_{j_k}(t,y)\prod_{\ell=1}^{\beta}\mathcal{Q}_{j_{\ell}}\left(\frac{1}{t},\frac{1}{y}\right)y^{\tau-\alpha-1}}{(y+1)^{\tau+1}(ty+1)^{\tau-\alpha-\beta-1}} {\rm d}y\\
\end{split}
\end{equation}
where \(K_m(\alpha,\beta)\) and \(\mathcal{Q}_{j_k}(t,y)\) \ are given by (\ref{eq:P(a,b,m)}) and (\ref{eq:S(t,y)}) respectively, at the top of this page. Further, we define 
\(\mathcal{U}_{i,k}(m,\alpha,\beta)=(m+k+1+j_k)_{i-1}(m+i-k-j_k)_{\alpha+\beta-i}\) and \(\mathcal{V}_{i,\ell}(m,\alpha,\beta)=(m+\ell-\alpha+1+j_{\ell})_{i-1}(m+i-\ell+\alpha-j_{\ell})_{\alpha+\beta-i}\)
with $\alpha=n-m$, $\beta=p-m $ and $\tau=(m+\alpha)(m+\beta)$. Here, the notation $(a)_k$  
represents the Pochhammer symbol. 
\begin{IEEEproof}
The proof is in Appendix A.
\end{IEEEproof}

\begin{theorem}\label{th:2}
Let \( \textbf{W}_1 \sim \mathcal{W}_m(p, \mu\textbf{I}_m+\gamma \mathbf{vv}^H) \) and \( \textbf{W}_2 \sim \mathcal{W}_m(n, \textbf{I}_m) \)
be two independent Wishart matrices. Then
the c.d.f. of \(\kappa(\bm{\widehat{\Psi}})\) of \(\bm{\widehat{\Psi}}= \mu^{-1}\textbf{W}_1 \textbf{W}_2^{-1} \) is given by
\vspace{-1.5mm}
\begin{equation}\label{eq:H1_integral}
F_{\kappa}(t;\mathcal{H}_1)\hspace{-1mm}
= \hspace{-1mm}\mathcal{J}_{m,n,p}(\gamma_{\!e})
\hspace{-1mm}\int_{\mathcal{D}(t)} \hspace{-2mm}
\big[\mathcal{P}^A_{m,n,p}(\gamma_{\!e}) + \mathcal{P}^B_{m,n,p}(\gamma_{\!e})\big]
\,{\rm d}\mathbf{z},
\end{equation}
where \(\gamma_{\!e}=\gamma/\mu\) and the integration domain is defined as
\[
\mathcal{D}(t) = \left\{\, \mathbf{z} = (y, \boldsymbol{\lambda}) 
\,\big|\, y \in [0, \infty), \; \boldsymbol{\lambda} \in [y, ty]^{m-1} \,\right\},
\]
and the composite differential is \( {\rm d}\mathbf{z} = {\rm d}\boldsymbol{\lambda}\,{\rm d}y \).
$\mathcal{J}_{m,n,p}(\gamma_{\!e})=\frac{\mathcal{K}_{m,n,p}}{\gamma_{\!e}^{m-1}(1+\gamma_{\!e})^{p+1-m}}$ with $\mathcal{K}_{m,n,p}$ defined in Appendix A. $\mathcal{P}^A_{m,n,p}(\gamma_{\!e})$ and $\mathcal{P}^B_{m,n,p}(\gamma_{\!e})$ are defined as
\begin{equation}\nonumber
    \mathcal{P}^A_{m,n,p}(\gamma_{\!e})=\frac{y^{p-m}}{\left( 1 + \frac{y}{\gamma_{\!e}+1} \right)^{\delta}} \prod_{j=2}^{m} \frac{\lambda_j^{p-m}(y-\lambda_j)}{(1+\lambda_j)^{p+n-1}}\Delta^2_{m-1}(\bm{\lambda}),
\end{equation}
\begin{equation}\nonumber
\begin{split}
    \mathcal{P}^B_{m,n,p}(\gamma_{\!e})=&\frac{(m-1)(\lambda_2-y)(y\lambda_2)^{p-m}}{(1+y)^{p+n-1}\left( 1 + \frac{\lambda_2}{\gamma_{\!e}+1} \right)^{\delta}}\\
    &\times\prod_{j=3}^m \frac{\lambda_j^{p-m}(y-\lambda_j)^2(\lambda_2-\lambda_j) \Delta^2_{m-2}(\bm{\lambda})}{(1+\lambda_j)^{p+n-1}}.
\end{split}
\end{equation}
\end{theorem} 
Here $\delta=p+n+1-m$. 
\begin{IEEEproof}
The result follows by applying the joint probability density function (p.d.f.) of the eigenvalues of the matrix \(\widehat{\bm{\Psi}}\) as given in~\cite[Eq. (13)]{Chamain}, followed by an integration procedure similar to that used in Appendix~A for evaluating the false alarm probability. Details are omitted for brevity.
\end{IEEEproof}

\begin{figure*}[t!]
    \centering
    \subfloat[ROC comparison and $t^*$ selection.\label{fig:ROC}]{\includegraphics[width=0.334\textwidth]{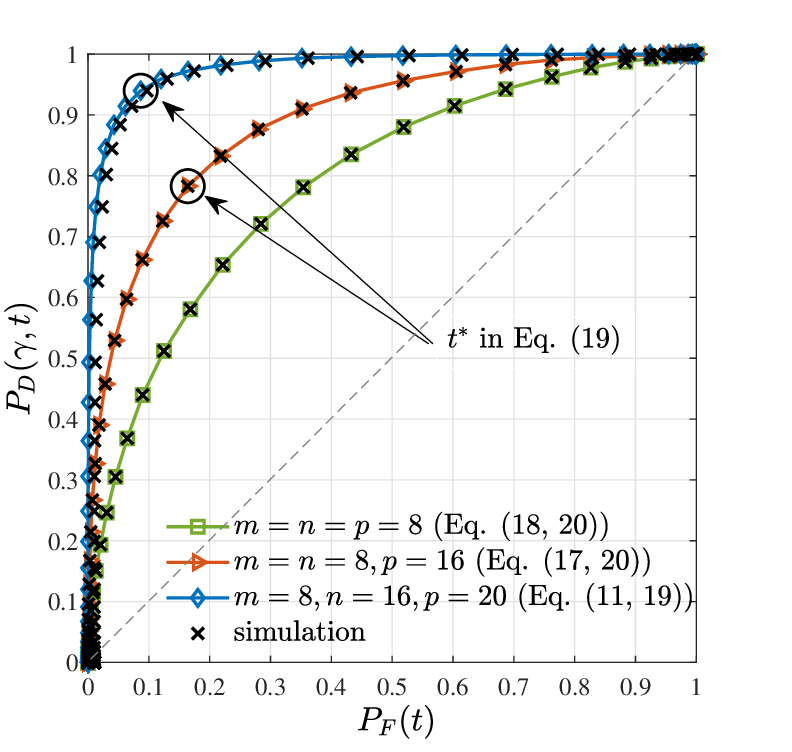}} 
    \hfill
    \subfloat[False-alarm probability vs. noise mismatch $\mu$.\label{fig:Pf_CFAR}]{\includegraphics[width=0.333\textwidth]{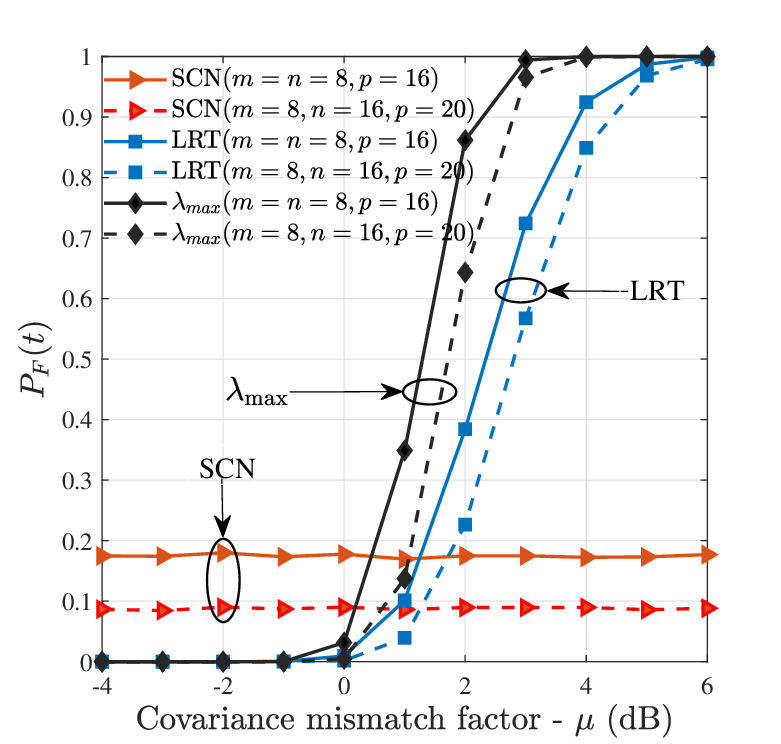}} 
    \hfill
    \subfloat[Total error $P_E(t)$ vs. noise mismatch $\mu$.\label{fig:total_error}]{\includegraphics[width=0.333\textwidth]{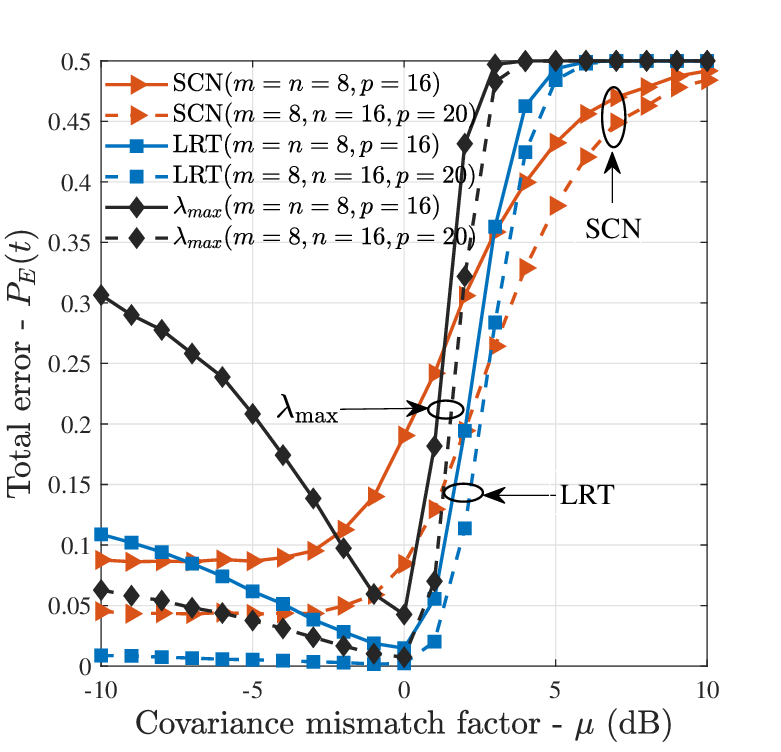}} 
    \caption{Performance comparison of SCN, LRT and $\lambda_{max}$: (a) ROC validation, (b) $P_{\mathrm{F}}$ behavior, and (c) total error comparison}
    \label{fig:simulation_results}\vspace{-5mm}
\end{figure*}

For a target false-alarm rate $\alpha$, the CFAR threshold is 
\begin{equation}
t_\alpha = F_{\kappa}^{-1}(1-\alpha;\mathcal{H}_0).
\end{equation}

\begin{remark}[CFAR and Effective SNR Scaling]
Under $\mathcal{H}_0$, $F_{\kappa}(t;\mathcal{H}_0)$ is invariant to the mismatch factor $\mu$, confirming that $P_{\mathrm{F}}(t)$ is CFAR.  
Under $\mathcal{H}_1$, the effective SNR reduces to
\(
\gamma_{\!e} = \gamma / \mu,
\)
so $F_{\kappa}(t;\mathcal{H}_1)$ depends on $(\gamma,\mu)$ only through $\gamma_{\!e}$.
\end{remark}

\begin{remark}[Optimal Threshold Selection]\label{remark_2}
For known SNR $\gamma$, the decision threshold $t$ can be optimized by minimizing the total error $P_{\mathrm{E}}(t)=P_{\mathrm{F}}(t) + (1 - P_{\mathrm{D}}(\gamma,t))$ as
\begin{equation}\label{eq:min_error_threshold}
t^* = \arg\min_t \left[P_{\mathrm{F}}(t) + (1 - P_{\mathrm{D}}(\gamma,t))\right],
\end{equation}
here $P_{\mathrm{F}}(t)$ and $P_{\mathrm{D}}(\gamma,t)$ are obtained from Theorems~\ref{th:1} and~\ref{th:2}.
\end{remark}

\begin{remark}[Closed-Form Special Cases]
Although complete closed-form expressions for $P_{\mathrm{F}}$ and $P_{\mathrm{D}}$ are analytically tractable, they are omitted for brevity.  
Simplified results are provided for representative cases: $P_{\mathrm{F}}$ when $m=n$ (Corollary~1.1) and both $P_{\mathrm{F}}$, $P_{\mathrm{D}}$ when $m=n=p$ (Corollaries~1.2 and~2.1).
\end{remark}

\textit{Corollary 1.1:} For $\alpha=0$ (i.e., $m=n$), the c.d.f. given in Theorem 1 degenerates to 
\begin{align}\label{eq:CDF_Alpha0}
&F_{\kappa}(t;\,\mathcal{H}_0)\nonumber\\ 
&=\widetilde{K}_m(\beta)\sum_{j_{1}=0}^{(m+\beta-2)}\sum_{j_{2}=0}^{(m+\beta-3)}\ldots\sum_{j_{\beta}=0}^{(m-1)}\prod_{\ell=1}^{\beta}\mathcal{G}_{\ell}(j_{\ell})\nonumber\\
& \times B\left(\nu,\,\sum_{\ell=1}^{\beta}j_{\ell}-\nu\hspace{-0.5mm}+\hspace{-0.5mm}\beta\right)(-t)^{\sum_{\ell=1}^{\beta}j_{\ell}}(t-1)^{\nu-\sum_{\ell=1}^{\beta}j_{\ell}-\beta-1}\nonumber\\
& \times{}_2F_1\left(\nu-\beta-1,\nu;\,2\nu-\beta-\sum_{\ell=1}^{\beta}j_{\ell};1-t\right),\; t>1
\end{align}
where ${}_2F_1(\cdot)$ is the Gauss hypergeometric function \cite{gradshteyn2007}, $B(\,,)$ denotes the Beta function. Here $\widetilde{K}_m(\beta)$ is defined in (\ref{eq:tildaQ}) with \(\mathcal{G}_{\ell}(p) = \frac{(m+\ell+p)!}{p!(\ell)_{p}[(\ell-1)!]^2(m+\beta-\ell-1-p)!}\) and $\nu=m(\beta+m)$.

{\bf Note}: This is valid under the constraint $(p-m)(p+m-1)<2mp$ leads to $m>\sqrt{0.5 \beta(\beta-1)}$, which ensures the convergence of the term-by-term integration in Theorem~1. However, such a restriction is not warranted when $\alpha$ and $\beta$ are zero simultaneously, as demonstrated in Corollary 1.2.

\textit{Corollary 1.2:} For $\alpha=0$ and $\beta=0$ (i.e., $m=n=p$), the c.d.f. given in Theorem 1 degenerates, for $t>1$, to 
\begin{equation}
\begin{split}
F_{\kappa}(t;\,\mathcal{H}_0) &= m^2\,\text{B}(m^2,m^2)\left(t-1\right)^{m^2-1}\\
& \quad \times {}_2F_1\left(m^2, m^2-1; 2 m^2; 1-t\right).\label{eq:H0AlphaBetaZero}
\end{split}
\end{equation}

\textit{Corollary 2.1:} For $\alpha=0$ and $\beta=0$ (i.e., $m=n=p$), the c.d.f. given in Theorem 2 degenerates, for $t>1$, to
\begin{align}\label{eq:H1_alphaBetaZero}
F_{\kappa}(t;\,\mathcal{H}_1) = \mathcal{I}^A_m(t) + \mathcal{I}^B_m(t),\;\; t>1
\end{align}
where $\mathcal{I}^A_m(t)$ and $\mathcal{I}^B_m(t)$ are defined by (\ref{eq:I_A}) and (\ref{eq:I_B}), respectively. Moreover, $\widetilde{m}=m^2-m+1$ and $F_1(\cdot)$ denotes the Appell hypergeometric function \cite{gradshteyn2007}.

\vspace{-2mm}

\section{Numerical Results}\label{s_per_reciprocal}
This section validates the unified analytical framework and assesses the robustness of the SCN detector under both ideal and disturbed sensing conditions.  
Fig.~\ref{fig:ROC} compares analytical and Monte Carlo results for representative $(m,n,p)$ configurations, showing near-perfect agreement with the derived CDF expressions in \eqref{eq:cdf_integral}–\eqref{eq:H1_alphaBetaZero}.  
Detection performance improves with increasing $m$, $n$, and $p$, consistent with the predicted asymptotic behavior from random matrix theory (RMT).

To evaluate robustness, thresholds are computed at $\mu=1$ using the total-error minimization rule (Remark~\ref{remark_2}) and reused for disturbed cases with $\mu\neq1$.  
The resulting false-alarm probability $P_{\mathrm{F}}$ and total error $\epsilon_{\mathrm{tot}}$ are then compared across SCN, $\lambda_{\max}$, and LRT detectors.  
As illustrated in Fig.~\ref{fig:Pf_CFAR}, the SCN preserves constant false-alarm across all $\mu$, confirming its CFAR behavior predicted by the analytical model, whereas $\lambda_{\max}$ and LRT show significant sensitivity to covariance scaling.  
Fig.~\ref{fig:total_error} further highlights that both LRT and $\lambda_{\max}$ experience steep error growth as $\mu$ deviates from $0$\,dB. In contrast, the SCN detector shows better resilience under high noise uncertainty, especially for $\mu > 4\,\text{dB}$, maintaining a lower \(P_E(t)\) than the other methods, highlighting its advantage in disturbed environments.

These findings verify that the proposed unified framework accurately predicts detector behavior under both stationary and non-stationary noise statistics.  
They also demonstrate that the SCN provides a consistent CFAR response and the lowest total error across extremely disturbed operating conditions, establishing it as a robust solution for interference-limited, dynamically varying sensing environments.

\vspace{-0.1cm}
\section{Conclusion}
\vspace{-0.1cm}
This paper developed a unified analytical framework for SCN-based signal detection in colored noise, covering both ideal (stationary) and disturbed (non-stationary) sensing conditions.  
By incorporating whitening from noise-only training samples, the framework models both consistent and mismatched noise covariances within a common formulation.  
Analytical expressions for false-alarm and detection probabilities were derived using random matrix theory, with closed-form solutions for tractable cases.  
The analysis demonstrated that the SCN statistic remains CFAR under covariance mismatch, whereas conventional detectors, such as the LRT and maximum-eigenvalue tests, lose robustness.  
Simulations validated the theoretical predictions, confirming consistent performance and low total error across varying interference levels.  
Overall, the SCN detector provides a mathematically rigorous and practically robust solution for spectrum sensing and ISAC applications under noise uncertainty.  
Future work will extend this framework to broader matrix structures and adaptive threshold optimization in dynamic environments.

\vspace{-0.1cm}
\section*{Appendix A (Proof of Theorem 1)}\label{sec:appendix_A}
\vspace{-0.1cm}
\begin{def}\label{def:1}
The joint density of the ordered eigenvalues \( \lambda_1 < \dots < \lambda_m \) of \( \mu^{-1}\mathbf{W}_1 \mathbf{W}_2^{-1} \), where \( \mathbf{W}_1 \sim \mathcal{W}_m(p, \mu\mathbf{I}_m) \) and \( \mathbf{W}_2 \sim \mathcal{W}_m(n, \mathbf{I}_m) \)  for \( p, n \geq m \), under \( \mathcal{H}_0 \), is given by
\end{def} 
\vspace{-2mm}
\begin{equation}
\begin{split}
f_0(\lambda_1, \ldots,\lambda_m)&=\mathcal{K}_{m,n,p}\prod_{j=1}^m\frac{\lambda_j^{p-m}}{(1+\lambda_j)^{p+m}}\Delta^2_m(\bm{\lambda}),
\end{split}
\end{equation}
where  
\mbox{$\Delta_m(\bm{\lambda}) = \prod_{1 \leq i < j \leq m} (\lambda_j - \lambda_i)$}, 
is the Vandermonde determinant and \mbox{\(\mathcal{K}_{m,n,p}=\frac{\pi^{m(m-1)}\widetilde{\Gamma}_m(n+p)}{\widetilde{\Gamma}_m(m)\widetilde{\Gamma}_m(n)\widetilde{\Gamma}_m(p)}\)} with \(\widetilde{\Gamma}_m(n)=\pi^{\frac{1}{2}m(m-1)}\prod_{j=1}^m\Gamma(n-j+1)\) where $\Gamma(\cdot)$ is the classical Gamma function.  
The c.d.f. of $\kappa(\widehat{\boldsymbol{\Psi}})$ under $\mathcal{H}_0$ by keeping the integration with respect to $\lambda_1=y$ last is given by
\begin{align}\label{eq:CDFH0H1Generalized}
   F_{\kappa}(t;\mathcal{H}_0)=
   \hspace{-2mm}\int_0^\infty \hspace{-1.5mm}\int _{\mathcal{S}}
f_0(y,\lambda_2,\ldots,\lambda_m) {\rm d}\lambda_2\ldots{\rm d}\lambda_m {\rm d}y
\end{align}
where $\mathcal{S}=\{y<\lambda_2<\ldots<\lambda_m<t\lambda_1\}$.
Due to the symmetry among terms \mbox{\(\lambda_2, \lambda_3 \ldots \lambda_m\)},
the region of integration for $\mathcal{H}_0$ can be rearranged as an unordered region to yield 
\begin{equation*}
\begin{split}
&F_{\kappa}(t;\mathcal{H}_0)=\frac{1}{(m-1)!}\int_0^\infty \int_{\tilde{\mathcal{S}}}f_0(y, \ldots,\lambda_m)\, {\rm d}\bm{\lambda} {\rm d}y\\
\end{split}
\end{equation*}
here $\tilde{\mathcal{S}}=[y,ty]^{m-1}=[y,ty]\otimes \ldots \otimes [y,ty] $ with $\otimes$ representing the Cartesian product and
${\rm d}\boldsymbol{\lambda}=\prod_{i=2}^m{\rm d}\lambda_i$.
It should be noted that the integration range differs for $y$ than the rest of the eigenvalues. Therefore, isolating the $y$ component from the rest and relabeling $\lambda_j=z_{j-1}$, \mbox{$j=2,3,\ldots m$},
and applying the variable transformations $u_j=z_j/(z_j+1)$, \mbox{$j=1,2,\ldots m-1$,} with some algebraic manipulation yields to
\begin{equation}\nonumber
\begin{split}
&F_{\kappa}(t;\mathcal{H}_0)
=\frac{\mathcal{K}_{m,n,p}}{(m-1)!}\int_0^\infty\frac{y^{\beta}}{(1+y)^{\alpha+\beta+2}} \\
&\times \int_{\tilde{\mathcal{S}}_{\ell}}\prod_{j=1}^{m-1}u_j^{\beta}(1-u_j)^{\alpha}\left({\frac{y}{(1+y)}-u_j}\right)^2\Delta_{m-1}^2(\bm{u}) {\rm d}\bm{u} {\rm d}y,
\end{split}
\end{equation}
where  $\tilde{\mathcal{S}}_{\ell}=[\ell_{\ell},\ell_u]^{m-1}$, \mbox{${\ell}_{\ell} = y/(y+1)$}, \mbox{${\ell}_u=ty/(ty+1)$}.

To evaluate this, we employ the orthogonal polynomial method from RMT~\cite{Chamain},  
mapping the domain to \((-1,1)\) and exploiting Jacobi polynomial orthogonality.  
This leads to the analytical c.d.f. of the SCN under \(\mathcal{H}_0\) in Theorem~\ref{th:1}.
\vspace{-1mm}



\end{document}